\documentclass[doublecol]{epl2} 

\newcommand{\x}{{\bm x}}

\newcommand{\be}{\begin{equation}}
\newcommand{\ee}{\end{equation}}
\usepackage{amsmath}

\title{Continuum Free-Energy formulation for a class of Lattice Boltzmann multiphase models}
\shorttitle{Free Energy formulation for a class of multiphase models} 

\author{M. SBRAGAGLIA\inst{1}, H. CHEN \inst{2} , X. SHAN \inst{2} AND S. SUCCI\inst{3}}
\shortauthor{M. Sbragaglia  \etal}

\institute{                    
  \inst{1} Department of Physics and INFN, University of Tor Vergata, Via della Ricerca Scientifica 1, 00133 Rome, Italy \\
\inst{2} EXA Corporation,  55 Network Drive, Burlington,  Massachussets 01803, USA \\
  \inst{3} Istituto per le Applicazioni del Calcolo CNR, Viale del Policlinico 137, 00161 Roma, Italy.}

\pacs{47.11.-j}{Computational Methods in Fluid Dynamics}
\pacs{47.45.-n}{Rarefied gas Dynamics}
\pacs{02.70.-c}{Computational Techniques; Simulations}

\abstract{It is shown that the Shan-Chen (SC) model for non-ideal lattice fluids can be made compliant with a pseudo free-energy principle by simple addition of a gradient force, whose expression is uniquely specified in terms of the fluid density.  This additional term is numerically shown to provide fairly negligible effects on the system evolution during phase-separation. To the best of our knowledge, these important properties of the SC model were not noted before. The approach developed in the present work is based on a continuum analysis: further extensions,  more in line with a discrete lattice theory (X. Shan, {\it Phys Rev E}, {\bf 77} 066702 (2008)) can  be envisaged for the future.}

\begin{document}

\maketitle

Multiphase fluid flows are ubiquitous in nature, science and technology \cite{Rowlinson,Brennen}.  
Owing to the complexity of the underlying physics, often leading to the formation of complex 
space-time changing interfaces between the various phases/components, analytical methods 
have a limited range of applicability for the quantitative study of multiphase flows. 
Consequently, flexible and efficient methods for the numerical simulation
of multiphase flows are in constant demand.  
In the last decade, a new class of methods, based on a suitably discretized version of the Boltzmann kinetic equation, i.e. the lattice Boltzmann (LB) models \cite{Gladrow,BSV,Chen}  have been developed \cite{SC1,SC2,Yeo,Pooley,Tlee,Bettolo,Wagner,Arcidiacono,CHEN}. 

The main appeal of the lattice kinetic approach is the simplicity of its underlying dynamics, which consists 
of a free-streaming of a set pseudo-particles along a set of prescribed directions (discrete speeds), followed 
by a collisional relaxation around local equilibria.  
Crucial to the extension of the LB scheme to non-ideal fluids, is the possibility of encoding potential energy interactions at the level of a simple forcing term, acting upon the discrete Boltzmann distributions. Suitable choices of this forcing term have proven capable of triggering a fairly rich non-ideal fluid behaviour, such as phase-transitions and interface formation/propagation
\cite{SC1,Hiva,PRLnostro}, with no need of tracking the shape of the interface, but rather having it emerge from the underlying mesoscopic interactions.

A widely used  lattice Boltzmann algorithm \cite{Gladrow} with interparticle interactions  is the pseudo-potential scheme due to Shan-Chen (hereafter SC) model \cite{SC1,SC2}. The evolution over a unit time lapse reads as follows:
\be\label{eq:LB}
f_{l}(\bm{x}+\bm{c}_{l},t+1)-f_l(\bm{x},t)=-\frac{1}{\tau}\left(f_{l}(\bm{x},t)-f_{l}^{(eq)}(\rho, {\bm u}^{\prime}) \right) 
\ee
where $f_l(\bm{x},t)$ is the kinetic probability density function associated with the particle velocity $\bm{c}_{l}$, $\tau$ is the mean collision time, $f^{(eq)}_{l}(\rho, {\bm u})$  the local equilibrium, corresponding to a Maxwellian distribution with local density $\rho$ and local fluid speed $\bm{u}$. Finally, $\rho {\bm u}^{\prime}=\rho {\bm u}+\tau {\bm F}$ where ${\bm F}$ represents a general forcing term encoding inter-molecular interactions.  To be noted that this force enters the scheme in the form of a shift, $\bm{u}^{\prime}-{\bm u} \equiv \bm{F} \tau/\rho$,  of the local flow speed. In the original SC model \cite{SC1,SC2}, the bulk interparticle interaction is proportional to a free parameter  (the ratio of potential to thermal energy), ${\cal G}$, entering the equation for the momentum balance: 
\be\label{forcing} 
F^{SC}_i({\bm x})={\cal G} \psi ({\bm x}) \sum_{l} w_{l}(|{\bm c}_l|^2) \psi({\bm x}+{\bm c}_{l}) {\bm c}_l^{i}.
\ee 
where $w_l(|{\bm c}_{l}|^2)$ are static weights normalized to unity \cite{Shan06}, and $\psi(\x,t)=\psi(\rho(\x,t))$ is the (pseudo) potential describing  the fluid-fluid interactions triggered by inhomogeneities of the density profile.  

This pseudo-potential may also be viewed as a {\it generalized density}, obeying the  general properties of going to zero in the limit $\rho \rightarrow 0$, and saturating 
to a constant value at large densities. 
The mapping $\rho \rightarrow \psi(\rho)$ may be heuristically interpreted as the result of  a resummation of many-body collisions at all orders. On practical grounds, its main effect is to set the intermolecular forces to zero  at high density, a property which we may pictorially call 'asymptotic freedom', since it indicates that at  high-density/short-distances, particles are set free from mutual interactions.  This property is crucial to prevent density collapse due to attractive interactions, and surrogates the effect of  hard-core repulsion, which would otherwise jeopardize the numerical stability of the scheme.  In fact, the replacement of physical density with a generalized is probably to be regarded as  a general constraint of lattice models of non-ideal hydrodynamics.


The presence of the pseudo-potential turns the bulk ideal equation of state $P_{b}(\rho)=\rho T$ (with $T$ the local temperature ) into a non-ideal one, $P_{b}(\rho)= \rho T - \frac{{\cal G}}{2} \psi^2$.   It is readily seen that a sufficiently large value of the coupling  parameter ${\cal G}$ leads to a Van der Waals-like loop, corresponding to the formation of a stable liquid-gas interface (phase-separation) \cite{SC1,SC2}.   The above concerns the behaviour of the bulk flow, sufficiently away from the interface.  In order to discuss interface physics, one needs to inspect the contribution of the interactions  to the non-ideal pressure tensor, defined by the relation:
\be\label{PT}
-\partial_{j} P_{i j}=F^{SC}_{i}-\partial_{i} (\rho T).
\ee
The properties of this tensor have been investigated in a number of recent publications \cite{PSEUDO,Shan}.  
It should be noted that a direct transcription of the force balance on a lattice with a finite set of velocities leads
to a discrete form of the pressure tensor \cite{Shan} that differs slightly from the continuum definition (\ref{PT}).  However, by suitably enlarging the set of discrete velocities \cite{PSEUDO}, $P_{ij}$ is well approximated \footnote{Technically speaking, this limit corresponds to the case $e_4 >>1$ in equations (22a) and (22b) and $\epsilon \approx 1$ in equation (25) of \cite{Shan}.} by its continuum limit
\be\label{PTSC}
P_{i j}=\left[P_b(\rho)-\frac{{\cal G}}{4} |{\nabla \psi}|^2-\frac{{\cal G}}{2}\psi \Delta \psi  \right]\delta_{i j}+\frac{{\cal G}}{2}\partial_{i} \psi \partial_{j} \psi. 
\ee
that satisfies (\ref{PTSC}) upon Taylor expansion of the forcing (\ref{forcing}). In the sequel, we shall refer to this expression of the pressure tensor, leaving the discussion on the  discrete lattice effects to a future work.\\

At variance with other LB formulations for non-ideal fluids, the SC was not derived from a free-energy functional, which has often been pointed out as a theoretical liability of the model. 

Yet, in view of its practical success, it is tempting to argue that there might be an 'hidden' pseudo-free energy (PFE) behind it.  
In this work, we shall show that the notion of generalized density can indeed be derived from a suitably generalized (pseudo) free-energy  functional.  
This is a potentially important result in two respects; first, it puts the SC method on a solid foundational basis within 
the general framework of non-equilibrium statistical mechanics.  
Second, it opens new perspectives for the formulation of a thermo-hydrodynamical lattice formulation of non-ideal fluids. 
A very convenient aspect of the SC model over direct free-energy formulations \cite{Yeo} is that 
local equilibria keep a Maxwellian shape, only shifted by the amount $\bm{F}\tau/\rho$.
This offers better control on its numerical stability, and also a direct link with self-consistent thermal equilibria \cite{LAST}.  

The former property is key for lattice implementations, since it permits to preserve the same "stream-collide" structure of LB schemes for ideal gases, the effects of non-ideal interactions being entirely incorporated within the self-consistent velocity shift. 
It is worth reminding that the stream-collide structure is essential to the efficiency/accuracy of the method, since the lattice streaming operator (lhs of equation (\ref{eq:LB})) is  {\em exact}, and collisions are conservative up to machine round-off. 

\section{Free-Energy derivation}

We begin by considering a free-energy functional in the standard weak-gradient approximation \cite{Rowlinson}:

\be\label{popular}
{\cal L}(\rho,{{\bm \nabla} \rho})=f(\rho)+\frac{\kappa}{2}|{\nabla \rho}|^2
\ee
where the bulk free energy $f(\rho)$ is connected to the bulk pressure via the familiar equilibrium Legendre's transform:
\be\label{Legendre}
P_b(\rho)=\rho \frac{d f}{d \rho}-f(\rho).
\ee
The term $\frac{\kappa}{2}|{\nabla \rho}|^2$ describes the energy surplus required to build up and maintain the interface, and is clearly related to surface tension \cite{Rowlinson}.  
Despite its manifest link to underlying molecular interactions\cite{Rochard68}, to the best of our knowledge,
no first-principle derivation of such relation is known. 
As a result, this term leaves us with some degree of latitude to design a free-energy functional 
consistent with the equilibrium states of the SC model. 

Since the off-diagonal terms of the SC pressure tensor in (\ref{PTSC}) take the form $ \approx \partial_i \psi \partial_j \psi$, it is natural to postulate the following {\it hybrid} ($\rho$-$\psi$) form of free-energy functional:   
$${\cal L}(\rho,{{\bm \nabla} \psi})=f(\rho)+\frac{\kappa}{2}   |{\nabla \psi}|^2=f(\rho)+\frac{\kappa}{2} \left(\frac{d \psi}{d \rho} \right)^2  |{\nabla \rho}|^2.$$
It is worth noting that this 'hybrid' form is equivalent to the standard one (\ref{popular}), only with a density-dependent 
surface coupling, $\kappa=\kappa(\rho) = \kappa \left(\frac{\partial \psi}{\partial \rho} \right)^2$.  This property has been recently exploited to produce a number of non-trivial dynamical
effects, such as structural arrest and ageing, in the framework of binary mixtures \cite{PRL09}. Having introduced the hybrid free-energy, we next proceed to identify the resulting conserved currents through a standard application of Noether's theorem.
 
Invariance of the free-energy with respect to spatial translations, leads to 
the following conserved tensorial current \cite{Goldstein}
\be
\partial_{j} J_{i j}=0
\ee
with 
\be
J_{i j}=-{\cal L} \delta_{i j}+\frac{\partial {\cal L}}{\partial (\partial_{i} \rho)}\partial_{j} \rho.
\ee
The next step is to explore the properties of this conserved current under the constraint of mass conservation 
and stationarity of the free energy. 
To this purpose, we introduce a Lagrange multiplier, $\lambda$, associated with mass conservation, and re-define  our PFE accordingly: 
\be\label{MASSCONS}
{\cal L}^{\lambda}(\rho,{{\bm \nabla} \psi})=f(\rho)+\frac{\kappa}{2} \left(\frac{d \psi}{d \rho}\right)^2  |{\nabla \rho}|^2-\lambda \rho.
\ee
Minimization of the constrained free-energy leads to the following Euler-Lagrange equations:
\be
\frac{\partial {\cal L}^{\lambda}}{\partial \rho}-\partial_{i} \left(\frac{\partial {\cal L}^{\lambda}}{\partial (\partial_{i} \rho)} \right)-\lambda=0
\ee
from which, the value of $\lambda$ is obtained:
\be
\lambda=\frac{d f(\rho)}{d \rho}-\kappa \frac{d \psi}{d \rho} \Delta \psi. 
\ee
With this value of $\lambda$ and the use of relation (\ref{Legendre}), the conserved current, hereafter to be identified with 
the free-energy consistent pressure tensor, ${P}^{FE}_{i j}$, becomes
\begin{equation}\label{tensor}
J_{i j}^{\lambda} \equiv {P}^{FE}_{i j}=\left[P_b(\rho)-\frac{\kappa}{2} |\nabla \psi|^2-\kappa\rho \frac{d \psi}{d \rho} \Delta \psi   \right]\delta_{i j}+\kappa\partial_{i} \psi \partial_{j} \psi
\end{equation}
where we observe that, as anticipated, in order to reproduce the off-diagonal term $\kappa\partial_{i} \psi \partial_{j} \psi$,  a contribution $\frac{\kappa}{2}|{\nabla \psi}|^2$ in the free energy is actually required.  This immediately fixes $\kappa=\frac{{\cal G}}{2}$.

\section{Connection to Shan-Chen Model}

The above derivation bears close links to the pressure tensor associated with the SC forcing, in its simplest instance:
$$F^{SC}_i({\bm x})={\cal G} \psi ({\bm x}) \sum_{l} w_{l}(|{\bm c}_l|^2) \psi({\bm x}+{\bm c}_{l} ) {\bm c}_l^{i}.$$
In particular, on the assumption of gentle interfaces, hence neglecting higher order terms, this force 
(we remind that in our notation ${\cal G}>0$) reads as:
\be\label{FORCE}
F^{SC}_{i}={\cal G} \partial_i \frac{\psi^2}{2}+\frac{1}{2}{\cal G} \psi \partial_i \Delta \psi+....  
\ee
where we have made use of the usual lattice tensor properties (see \cite{PSEUDO}  for a detailed description). 
The first term in (\ref{FORCE}) is related to the bulk pressure via the relation (\ref{PT}), 
$$P_b(\rho)=\rho T-\frac{1}{2} {\cal G} \psi^2(\rho).$$ 
The required function  $f(\rho)$ is readily derived from its definition as a Legendre's transform of the bulk pressure
$$f(\rho)= \rho \int_{\rho_0}^{\rho} \frac{P_b(\xi)}{\xi^2} d \xi  $$
where the constant of integration $\rho_0$  affects the total free energy through an additive constant, with no effect 
on the time derivative, because mass is assumed to be conserved.  
The surface term $\frac{1}{2}{\cal G} \psi \partial_{i} \Delta \psi$ has to be handled with care.  
The forcing term associated with the consistent free-energy pressure tensor in the continuum, reads 
(after setting $\kappa={\cal G}/2$ in (\ref{PTSC})) as follows:
\be
{P}^{FE}_{ij}=\left[P_b(\rho)-  \frac{{\cal G}}{4} |{\nabla \psi}|^2- \frac{{\cal G}}{2} \rho \frac{d \psi}{d \rho} \Delta \psi    \right]\delta_{i j}+\frac{{\cal G}}{2}\partial_{i} \psi \partial_{j} \psi.
\ee
This does {\it not} match the SC formulation given in (\ref{PTSC}), due to the presence of the  term $\frac{{\cal G}}{2} \rho \frac{d \psi}{d \rho} \Delta \psi$.  
The reason is clear: physical mass, associated with $\rho$, is conserved, while generalized mass, associated with $\psi$, is not.
This is why, strictly speaking, the SC model cannot be derived from a free-energy functional. 
Nevertheless, the interesting observation is that the only difference is confined to the diagonal component of the pressure tensor, while 
the off-diagonal term, $\frac{{\cal G}}{2}\partial_{i} \psi \partial_{j} \psi$, is exactly reproduced.  
This is a very welcome property, since it allows to match the two descriptions by simply adding an ad-hoc correction in the form of the gradient of a -known- potential.  
More precisely, we start from the following relation:
\be\label{relation}
{P}^{FE}_{ij}={P}^{SC}_{ij}+P_{ij}^{(add)}
\ee
where we have defined: 
\be
P^{(add)}_{ij}=\frac{{\cal G}}{2} \left[\zeta(\rho) \Delta \psi  \right] \delta_{ij}.
\ee
In the above,
\be
\zeta(\rho)=-\rho \frac{d \psi}{d \rho} +\psi
\ee
is the Legendre's transform of the generalized versus physical density. 
This term is responsible for the non-conservation of the generalized mass. 
Upon differentiation of expression (\ref{relation}) and use of (\ref{PT}), we obtain
\be
-\partial_{j} {P}^{FE}_{ij}+\partial_{i} (\rho )=F^{SC}_{i}-\partial_{i}  {V}^{FE} (\rho,\Delta \psi) 
\ee
where we have defined
\be
{V}^{FE} (\rho,\Delta \psi) = \frac{{\cal G}}{2} \left[\zeta(\rho) \Delta \psi  \right].
\ee  
Clearly ${V}^{FE}(\rho,\Delta \psi)$  is the precise free-energy potential needed to recover a consistent free-energy treatment. 
In other words, 
\be
{F}^{FE}_{i}=F^{SC}_{i}-\partial_{i} {V}^{FE}
\ee
is the right forcing, stemming from a free-energy consistent equilibrium.  In conclusion, the configuration minimizing the free-energy functional, whose density is
\be\label{FINALSTATE}
{\cal L}(\rho,{{\bm \nabla} \psi})=  f(\rho)+\frac{{\cal G}}{4} |{\nabla \psi}|^2 \;, \hspace{.2in} f(\rho)=\rho \int_{\rho_0}^{\rho} \frac{P_b(\xi)}{\xi^2} d \xi
\ee
and the static equilibrium associated with the forcing:
\be\label{PFEforcing}
{F}^{FE}_{i}({\bm x})={\cal G} \psi ({\bm x}) \sum_{l} w_{l}(|{\bm c}_l|^2) \psi ({\bm x}+{\bm c}_{l} ) {\bm c}_l^{i} -\partial_{i} {V}^{FE}
\ee
share the same physics.

\section{Continuum Maxwell Rule}

Next, we proceed to show that the integral constraints associated with the continuum free-energy density, lead 
to the Maxwell equal-area rule fixing the liquid and gas density ($\rho_l$ and $\rho_g$), independently of the 
structure of the interface, i.e. independently on whether the physical or generalized densities are used. 
This is again crucial to the practical success of the SC model.
To this end, we begin by minimizing ${\cal L}^{\lambda}(\rho,{{\bm \nabla} \psi})$ in (\ref{MASSCONS}) 
along a flat and stable liquid-gas interface, where inhomogeneities are confined along the $y$ coordinate. 
This sets the value of $\lambda$
\be\label{lambdanow}
\lambda=\frac{d f(\rho)}{d \rho}-\kappa \frac{d \psi}{d \rho} \partial_{yy} \psi 
\ee
with the bulk phases at the same chemical potential (because $\frac{d \psi}{d \rho} \partial_{yy} \psi=0$ in the bulk)
\be
\left. \frac{d f(\rho)}{d \rho} \right |_{\rho_g}=\left.  \frac{d f(\rho)}{d \rho} \right |_{\rho_l}
\ee
Upon multiplying by $\frac{d \rho}{d y}$ and integrating between $y=0$ (conventionally, the bulk gas phase) and 
a generic $y$, at density $\rho(y)$, we obtain:
\be
\frac{d f}{d y}-\kappa\frac{d \psi}{d y} \frac{d^2 \psi}{d y^2}=\lambda \frac{d \rho}{d y}.
\ee
Integrating once more along $y$, and making use of the relation (\ref{lambdanow}), leads to the following identity
\be
P_b(\rho)-P_b(\rho_g)=\frac{\kappa}{2} \rho^2 \frac{d}{ d \rho} \left[\left( \frac{d \psi}{d y} \right)^2 \frac{1}{\rho}  \right].
\ee
By calling $P_b(\rho_g)=P_0$ the constant value we wish to achieve, we finally obtain
\be
\int_{\rho_g}^{\rho} (P_{b}(\rho)-P_0)\frac{d \rho}{\rho^2}= \frac{\kappa}{2}  \left( \frac{d \psi}{d y} \right)^2 \frac{1}{\rho}.  
\ee
Since the rhs is zero in the bulk phase, we finally conclude that 
\be
\int_{\rho_g}^{\rho_l} (P_{b}(\rho)-P_0)\frac{d \rho}{\rho^2} = 0
\ee
which is the desired expression of Maxwell's equal-area law, q.e.d.
\section{Numerical tests}
In this section, our theoretical findings are tested against numerical simulations.  For the sake of simplicity, we shall refer to phase-separation in a $2d$ system at a constant temperature ($T=1$), as explained in \cite{Shan06}.

\begin{figure*}[t!]
\begin{minipage}[t!]{16cm}

\includegraphics[width=16cm,keepaspectratio]{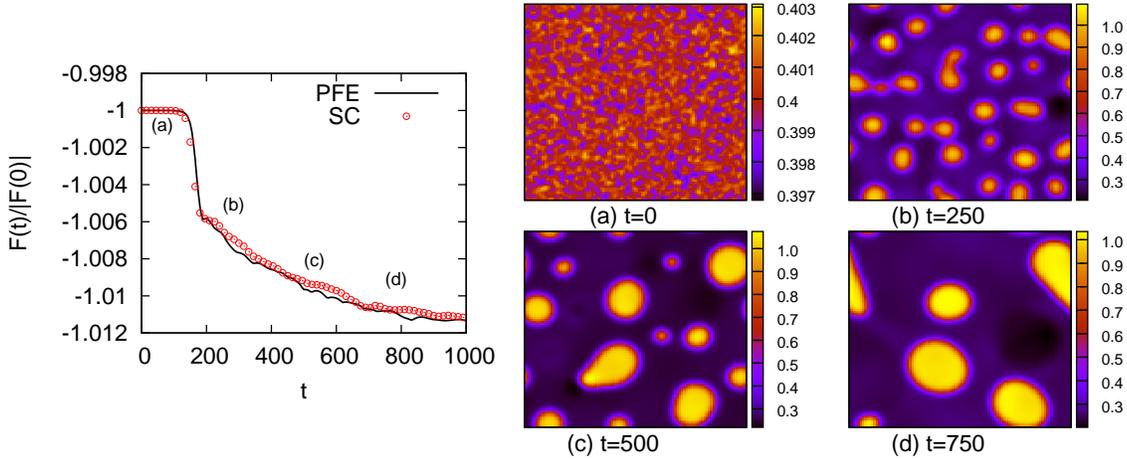}
\end{minipage}
\hfill
\begin{minipage}[t!]{16cm}
\caption{\label{fig1}Numerical simulation of two-dimensional phase-separation using the SC model.
Separation proceeds from an initial (see plot (a)) random configuration. 
We monitor the behaviour of the total PFE $F(t)=\int  {\cal L}(\rho,\nabla \psi) dx dy$ with ${\cal L}$ given in equations (\ref{FREENUM}) and (\ref{FINALSTATE}). The total PFE is normalized to its initial absolute value. 
A sequence of density contour plots are displayed in (a)-(d), corresponding to $t=0,250,500,750$ time units. 
In the left panel, we compare two characteristic trajectories of the total PFE, as obtained with a PFE consistent 
forcing (\ref{PFEforcing}) and with a bare SC forcing (\ref{forcing}), with
${\cal G}=4.3$ and $\psi(\rho)=e^{-\frac{1}{2 \rho}}$. 
In both cases we have used the lattice Boltzmann equation (\ref{eq:LB}) with $\tau=0.7$.}
\end{minipage}
\end{figure*}
We also have chosen\footnote{with this choice, the integral for $f(\rho)$ in (\ref{FINALSTATE}) can be computed exactly}, 
the pseudo-potential in the form $\psi(\rho)=e^{-\frac{1}{2 \rho}}$, which sets the free-energy density to 
\be\label{FREENUM}
f(\rho)=\rho T \log \rho - \frac{{\cal G}}{2} \rho \psi^2 
\ee
where the first term is the ideal gas contribution, supplemented by an interaction term. 
The value of ${\cal G}$ has been chosen equal to ${\cal G}=4.3$, so as to produce stable interfaces 
with densities $\rho_g=0.26$ (gas phase) and $\rho_l=1.03$ (liquid phase) approximately with an interface of the order of ten grid points. 

\begin{figure*}[t!]
\begin{minipage}[t!]{16cm}

\includegraphics[width=16cm,keepaspectratio]{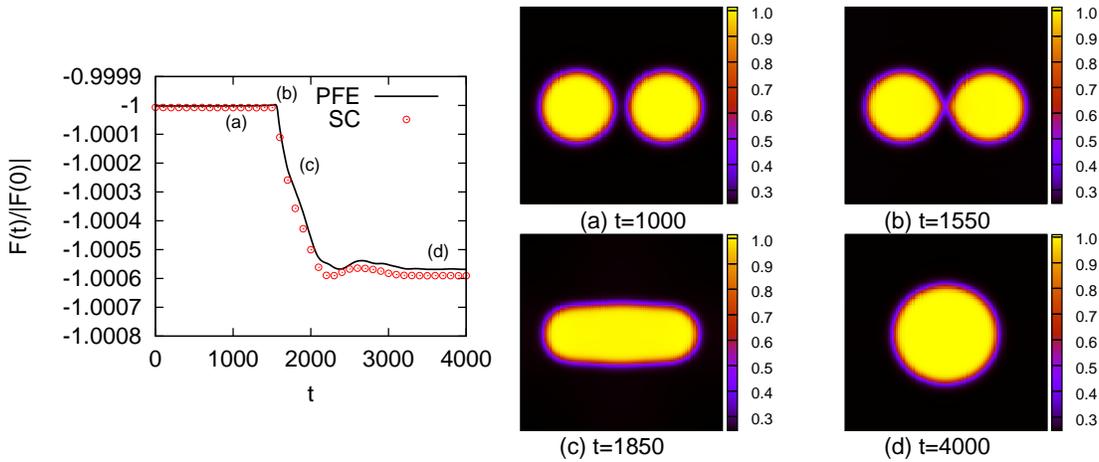}
\end{minipage}
\hfill
\begin{minipage}[t!]{16cm}
\caption{\label{fig2}Numerical simulation of two-dimensional droplet merging with the SC model. 
We monitor the behaviour of the total PFE, as computed from $F(t)=\int  {\cal L}(\rho,\nabla \psi) dx dy$ with ${\cal L}$ given in equations (\ref{FREENUM}) and (\ref{FINALSTATE}). As in figure \ref{fig1}, we compare two characteristic behaviours 
obtained with a PFE-consistent forcing (\ref{PFEforcing}) and with a bare SC forcing (\ref{forcing}), both with ${\cal G}=4.3$ and a pseudopotential given by $\psi(\rho)=e^{-\frac{1}{2 \rho}}$. 
In both cases we have used the lattice Boltzmann equation (\ref{eq:LB}) with $\tau=0.7$.}
\end{minipage}
\end{figure*}

In figure \ref{fig1} we show the time evolution of the total pseudo-free energy
$F(t)=\int {\cal L}(\rho,\nabla \psi) dx dy$, under the SC dynamics. 
We have used periodic boundary conditions and an initial state with small random perturbation around an average density 
value (see plot $(a)$ of this figure).  
The PFE shows a clear monotonic decrease in time, indicating that the SC dynamics does indeed
support some minimum principle. 
This finding is further corroborated by a direct comparison of the SC dynamics, as given 
by equation (\ref{forcing}), against the PFE dynamics, as obtained from the corrected model (\ref{PFEforcing}). 
The numerical data clear indicate that the lack of the correction term, not only does not spoil the  monotonic decrease of the PFE, but also provides a {\it quantitative} agreement with the correct PFE dynamics. We conclude that, notwithstanding its conceptual importance, the term restoring compliance of the SC model with a continuum free-energy functional, is fairly 
negligible on practical grounds. 
This is the basic reason behind the success of the SC model, and one that clearly hinges
on the fact that the departure between physical and generalized densities, as measured by the 'metric' $\zeta(\rho)$, remains sufficiently "small" all along the evolution.  The picture is further clarified by monitoring the detailed dynamics of  an 'elementary' coalescence event, i.e.  two droplets merging into a single one (equilibrium state). 
From the results reported in figure \ref{fig2}, the trend to a minimum PFE is again clearly visible, if  only with a much smaller change of the ratio $F(t)/F(0)$, as compared to the phase-separation test, which was clearly involving several merging events. In particular, the figure reveals that virtually all of the change in PFE is achieved during the merging event. 
Again, close agreement between the SC versus PFE dynamics is observed, indicating that the quantitative match observed in the phase-separation test applies to each merger event  separately, and is not due to some form of statistical cancellation between different merging events.

\section{Conclusions and outlook}

Summarizing, we have presented a number of theoretical findings on the properties of the SC model for non-ideal fluids.  
In particular, we have shown that the SC model supports a genuine pseudo-free-energy, monotonically decreasing in time as the fluid phase-separates. According to the common tenet, the SC dynamics does {\it not} derive from a continuum free-energy  functional; however, 
we have shown that it can be made compliant with a free-energy functional, by simply adding a gradient force, whose 
expression is uniquely specified in terms of the density field. 
Numerical simulations of a phase-separating fluid, provide a clear indication that
the additional gradient force plays a fairly negligible role on the evolution of the system. 
This unveils one of the main reasons behind the practical success of the SC method; although it was not derived  from a free energy minimization principle, the SC dynamics does possess a pseudo-free energy nonetheless.  To the best of our knowledge, these crucial properties of the SC model were not noticed before.

The list of the good news completed, a few words of caution are in order.
First, we have related the SC dynamics to a continuum free energy functional.  
While this is in order as a zero-th approximation, the 'discrete' limit of this analysis needs to be inspected very carefully \cite{Shan}. 
In particular, one would like to relate the SC dynamics to a {\it discrete} free-energy, i.e. a free-energy functional constructed directly on the lattice. The inclusion of (lattice) free energies and, more generally, interaction energies,  faces with the important 
issue of the existence of a H-theorem for lattice models with non-ideal interactions,  as well
as with the correct formulation of lattice thermo-hydrodynamic models \cite{LAST}.  The investigation of lattice models with non-ideal interactions and thermal dynamics, represents  a major issue for future research in the field, with a potential far-reaching impact for non-equilibrium statistical mechanics \cite{LEBOW}(flowing systems far from equilibrium).

\acknowledgments

We gratefully acknowledge discussions with R. Benzi, L. Biferale and G. De Prisco.

%
%

%

\acknowledgments

\end{document}